\documentclass[12pt,a4paper,final]{iopart}

\usepackage{iopams}
\usepackage{graphicx}
\usepackage[breaklinks=true,colorlinks=true,linkcolor=blue,urlcolor=blue,citecolor=blue]{hyperref}

\begin{document}

\title[Two-gap superconducting properties of intercalated iron-selenides]{Two-gap superconducting properties of alkaline-earth intercalated $A_{x}(NH_{3})Fe_{2}Se_{2}$ (A = Ba or Sr)}

\author[cor1]{Yung-Yuan Hsu}
\address{Department of Physics, National Taiwan Normal University, Taipei 11677, Taiwan, R.O.C.}
\ead{\mailto{yungyuan.hsu@gmail.com}}

\author{Yu-Bo Li}
\address{Department of Physics, National Taiwan Normal University, Taipei 11677, Taiwan, R.O.C.}
\ead{njvulfu761218@yahoo.com.tw}

\author{Shou-Ting Jian}
\address{Department of Physics, National Taiwan Normal University, Taipei 11677, Taiwan, R.O.C.}
\ead{drizzle12657387@hotmail.com}

\author{Gu-Kuei Li}
\address{Department of Physics, National Taiwan Normal University, Taipei 11677, Taiwan, R.O.C.}
\ead{tracephysics@hotmail.com}

\author{Ming-Cheng Yang}
\address{Department of Physics, National Taiwan Normal University, Taipei 11677, Taiwan, R.O.C.}
\ead{hulumiow@gmail.com}

\begin{abstract}
Superconducting properties were studied on high quality superconductors $Ba_{x}(NH_{3})Fe_{2}Se_{2}$ ($T_{c}$ = 39 K) and $Sr_{x}(NH_{3})Fe_{2}Se_{2}$ ($T_{c}$ = 44 K) prepared by intercalating Ba/Sr atoms into tetragonal $\beta$-FeSe by liquid ammonia. The elongated c-axis and almost unchanged a-axis of $Ba_{x}(NH_{3})Fe_{2}Se_{2}$, comparing with $\beta$-FeSe, suggested an unchanged intra-$Fe_{2}Se_{2}$-layer structure and the $T_{c}$ enhancement is due to a 3D to 2D-like Fermi surface transformation. The superconducting coherent lengths $\xi$(0), Ginzburg-Landau parameters $\kappa$ and penetration depths $\lambda$(0) obtained from the extrapolated lower and upper critical fields $B_{c1}$(0) and $B_{c2}$(0) indicates that both compounds are typical type-II superconductors. The temperature dependence of 1/$\lambda^{2}$(T) of $Ba_{x}(NH_{3})Fe_{2}Se_{2}$ deduced from the low field magnetic susceptibility shows a two-gap s-wave behaviour with superconducting gaps of $\Delta_{1}$ = 6.47 meV and $\Delta_{2}$ = 1.06 meV.

\end{abstract}

%\pacs{74.25.Dw, 74.25.Ha, 74.70.-b}

%\submitto{\SUST}
%\maketitle
\section{Introduction}

Since the discovery of iron-based superconductors in 2008~\cite{Kamihara2008}, the $T_{c} \simeq$ 8 K superconducting $\beta$-FeSe~\cite{McQueen2009} has attracted much interest duo to its simple lattice structure and sharing a common electronic origin for superconducting mechanism with the much complicated iron-arsenide systems~\cite{Greene2010,Kordyuk2012}. The $T_{c}$ of iron-selenide can be raised upto 14.5 K by tellurium partial substitution on selenium site~\cite{Yeh2008}. A break through of the iron-selenides system was achieved by the discovery of superconductors $A_{1-x}Fe_{2-y}Se_{2}$ (A = K, Rb, Cs or Tl, FeSe-122)~\cite{Guo2010,Ye2011} for their high $T_{c} \simeq$ 32 K and being a direct comparison system in electronic structure with iron-arsenide 122 systems (FeAs-122) and relative pnictide systems~\cite{Wang2012,Shoemaker2012,Li2012}. However, the unavoidable nano-scaled coexistence of antiferromagnetic insulated 245-phases ($T_{N}$ = 470-560 K) inside the superconducting FeSe-122 phase interferes the discussions of fundamental superconducting and electronic properties~\cite{Wang2012,Shoemaker2012,Li2012,Chen2011,Ksenofontov2011,Yuan2012}. 

Recently, an exciting development about pure superconductors with $T_{c}$ up to 46 K were reported for metal intercalated iron-selenides by ammonothermal reaction~\cite{Ying2012,Burrard2013}. The neutron diffraction studies indicated that ammonia molecules were inserted together with lithium atoms in-between the $Fe_{2}Se_{2}$-layers in the $T_{c}$ = 43 K $Li_{x}(NH_{3})Fe_{2}Se_{2}$ superconductor~\cite{Burrard2013}. It has been found that $T_{c}$ of the iron-selenides systems increases with an increasing adjacent $Fe_{2}Se_{2}$-layers distance~\cite{Zhang2013}. On the other hand, structural variations of intra-$Fe_{2}Se_{2}$-layer and electron doping as the origin of $T_{c}$ enhancement after the Li and $NH_{3}$ intercalation is also reported.~\cite{Burrard2013,Subedi2008}. 

In this report, structural and superconducting properties of $A_{x}(NH_{3})Fe_{2}Se_{2}$ (A = Ba or Sr) were studied. The structural analysis suggested that the intra-$Fe_{2}Se_{2}$-layer structure of $Ba_{x}(NH_{3})Fe_{2}Se_{2}$ remains the same as the parent superconducting FeSe compound. The enhancement of $T_{c}$ could be attributed to the more 2D-like electronic structure by shortened Brillouin zone z-axis due to inter-$Fe_{2}Se_{2}$-layer intercalation. Superconducting parameters, coherence length $\xi(0)$, penetration depth $\lambda(0)$, and Ginzburg-Landau parameter $\kappa$ extracted from magnetic measurements indicates both systems are typical type-II superconductors. Furthermore, superfluid density $n_{s} \approx \lambda^{-2}(T)$ of $Ba_{x}(NH_{3})Fe_{2}Se_{2}$ is found to be well described by a two-gap s-wave model similar to that of the parent compound~\cite{Khasanov2008}. 

\section{Experimental}

High quality powder samples of $A_{x}(NH_{3})Fe_{2}Se_{2}$ (A = Ba, Sr) were synthesized by intercalating Ba (or Sr) atoms into tetragonal $\beta$-FeSe using liquid ammonia (LA)~\cite{Ying2012,Burrard2013}. First, high-purity superconducting $\beta$-FeSe was prepared by high temperature reactions~\cite{McQueen2009}. Iron granules (99.98\%) and selenium shots (99.999\%) with 1.008:1 molar ratio were placed in an alumina crucible and sealed in a quartz tube. The tube was slowly heated to 750 $^{\circ}$C and held 1 day for complete reaction, then melted by heating to 1080 $^{\circ}$C followed by quenching to 420 $^{\circ}$C and held for 2 days for pure $\beta$ phase. The intercalating reaction was carried out by placing $\beta$-FeSe powder with Ba (99.7\%) or Sr (99\%) metal in a 4:1 molar ratio in an evacuated autoclave cooled in a liquid nitrogen bath~\cite{Ying2012}. Gaseous ammonia was slowly condensed into liquid until a Ba/Sr in LA concentration of 0.2-0.3 at$\%$ was reached. The vessel was kept at room temperature and magnetically stirred for 3 days. The obtained fine sample powders were then pressed into pellets and encapsulated by epoxy to prevent sample degradation due to ammonia escape. The lattice structure analysis was carried out by X-ray diffraction with a PHILIPS X$\rq$PERT diffractometer for 2$\theta$ range of 5-60 degree. The magnetic measurements were carried out by a QUANTUM DESIGN MPMS2 SQUID magnetometer with temperature down to 5 K and applied magnetic field up to 1 T.

\section{Results and discussions}

\begin{figure}[htb!]
 \includegraphics[width=70mm]{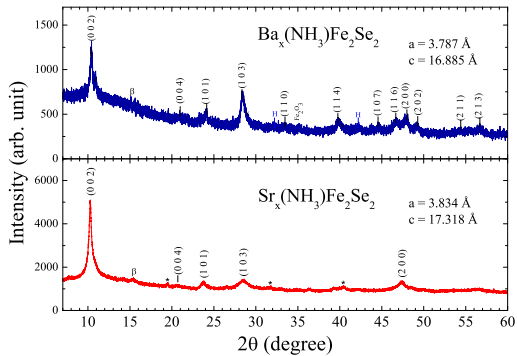}
 \caption{\label{fig.1}Powder X-ray diffraction patterns of $Ba_{x}(NH_{3})Fe_{2}Se_{2}$ (upper panel) and $Sr_{x}(NH_{3})Fe_{2}Se_{2}$(lower panel). The small peaks corresponding to hexagonal $\delta$-FeSe are labeled by "H", iron oxide by "$Fe_{2}O_{3}$" and unknown phases by asteroids.}
\end{figure}

The powder X-ray diffraction patterns of $A_{x}(NH_{3})Fe_{2}Se_{2}$ (A = Ba or Sr),  as shown in Figure 1,  can be well indexed by body-centered-tetragonal (bct) $Li_{0.6}(NH_{3})Fe_{2}Se_{2}$-type structure (space group: I4/mmm)~\cite{Burrard2013}. Minor phases of impurities were barely observed for hexagonal FeSe (marked by ``H" and iron oxide (by ``$Fe_{2}O_{3}$"), and unknown phases (by asteroids). No trace of superconducting precursor $\beta$-FeSe was observed. The derived lattice parameters for A = Ba are a = 0.3787 nm and c = 1.6885 nm, and for A = Sr are a = 0.3834 nm and c = 1.7318 nm. The greatly elongated c-axes observed are consistent with the literatures and can be attributed to $NH_{3}$ molecules co-intercalation~\cite{Ying2012,Burrard2013,Ying2013,Zalkins2008}. Lager lattice parameters and unit cell volume observed for Sr-compound (V = 0.2542 nm$^{3}$) than Ba-compound (V = 0.2422 nm$^{3}$) suggests a higher content of intercalated metal atoms in the former, which is consistent with the preliminary refinement results that the stoichiometric parameter x of $\sim$0.25 and $\sim$0.4 for A = Ba and Sr, respectively. However, due to the signal to noise ratio of the XRD data is not conclusive enough, we left the Ba/Sr content as undetermined in the following discussion. 

\begin{figure}[htb!]
 \includegraphics[width=70mm]{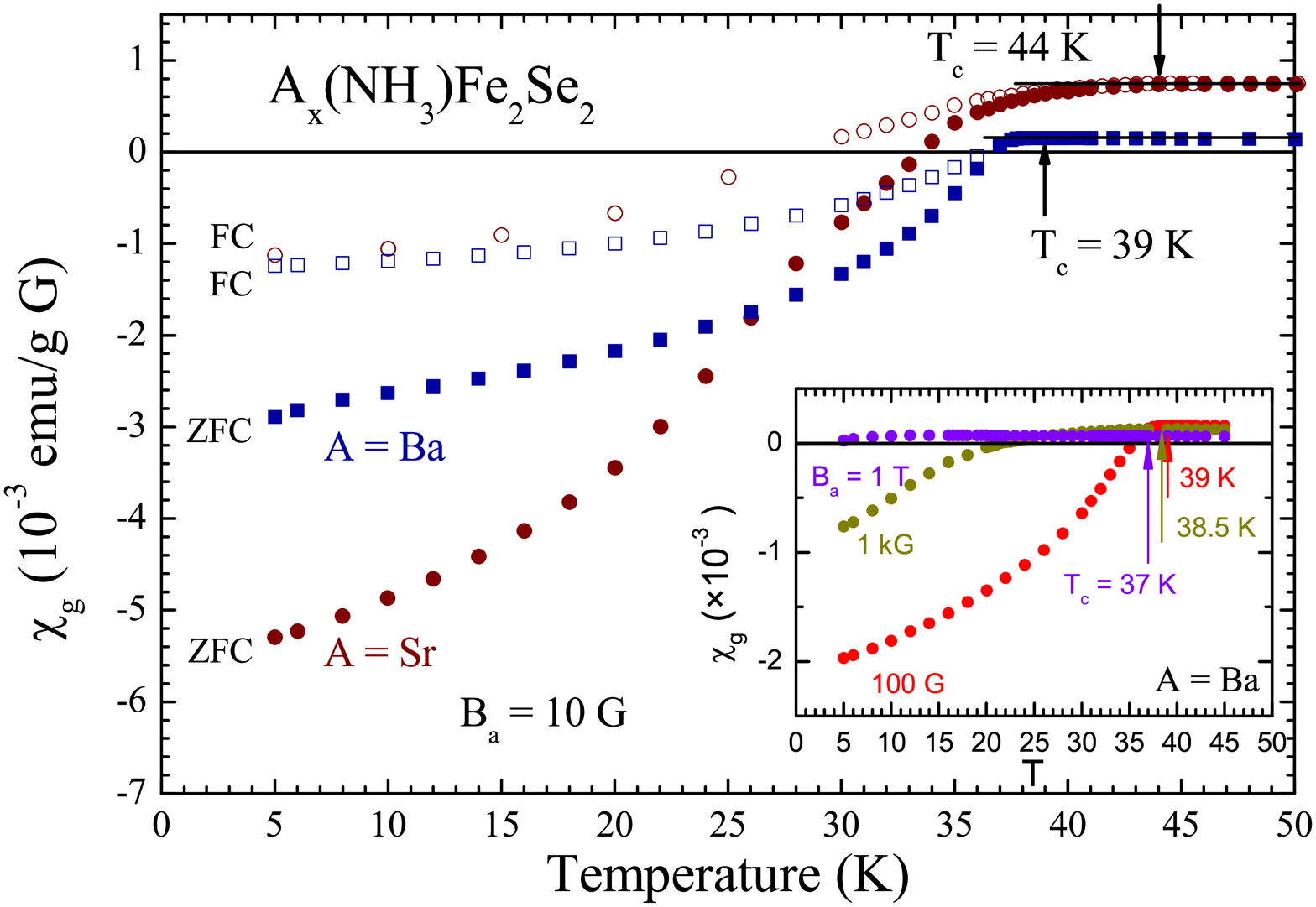}
 \caption{\label{fig.2}
 Low apply field ($B_{a}$ = 10 G) diamagnetic mass susceptibility 
$\chi_{g}(T)$ of superconducting $Ba_{x}(NH_{3})Fe_{2}Se_{2}$ and $Sr_{x}(NH_{3})Fe_{2}Se_{2}$ in zero-field-cooled (ZFC, solid symbols) and field-cooled (FC, open symbols) modes. An enhancement of diamagnetism below 14 K is attributed to the opening of smaller superconducting gap in the two-gap scenario. (Inset) Higher field (ZFC) $\chi_{g}-T$ curves of $Ba_{x}(NH_{3})Fe_{2}Se_{2}$ were used for $B_{c2}(T)$ determination.}
\end{figure}

The observed superconducting transition temperature $T_{c}$ for $Ba_{x}(NH_{3})Fe_{2}Se_{2}$ is 39 K and for $Sr_{x}(NH_{3})Fe_{2}Se_{2}$ is 44 K, as shown in Figure 2. A relatively large paramagnetic background was observed in the normal state, which could be attributed to the minor magnetic phases observed in XRD patterns. An enhancement of diamagnetism below 14 K can be clearly observed in $Ba_{x}(NH_{3})Fe_{2}Se_{2}$. This phenomenon is attributed to the opening of second superconducting gap in the weak-coupling two-gap scenario.  Similar behaviour was barely seen for Sr-compound due to its much larger particle sizes of the powder.

Among all intercalated iron-selenides superconductors, the enhancement of $T_{c}$ to around 40 K is believed due to the c-axis expansion and carrier density tuning by atoms/molecules intercalation~\cite{McQueen2009,Burrard2013,Ying2013}. However, the almost unchanged a-axis length and very low content of inserted Ba in $Ba_{x}(NH_{3})Fe_{2}Se_{2}$ suggests that one can consider the $Ba_{x}(NH_{3})Fe_{2}Se_{2}$ as directly separating $Fe_{2}Se_{2}$ layers in $\beta-FeSe$ by the inserted atoms/molecules without changing its intra-layer structure. This insertion elongates the c-axis length which weakens the inter-$Fe_{2}Se_{2}$-layer coupling. Consequently, it shortens the corresponding Brillouin zone c*-axis length in the reciprocal space, which makes the Fermi surfaces more cylindrical and more two-dimensional-like assuming the band structure is mainly determined by the almost unchanged intra-$Fe_{2}Se_{2}$-layer structure~\cite{Subedi2008}.

Inset of figure 2 shows the higher-field $\chi_{g}$ for $Ba_{x}(NH_{3})Fe_{2}Se_{2}$ which were measured right after the 10-G measurement to avoid any sample degradation problems.  The observed $T_{c}$ determined by the deviation point from the normal state decreases slightly to 37 K for $B_{a}$ = 1 T. The magnitude of diamagnetic signal decreases rapidly with increasing applied field. The ZFC $\chi_{g}$(1 T) curve even did not change sign down to 5 K due to fine powder grand size of the sample powder and the strong paramagnetic background.

\begin{figure}[htb!]
 \includegraphics[width=70mm]{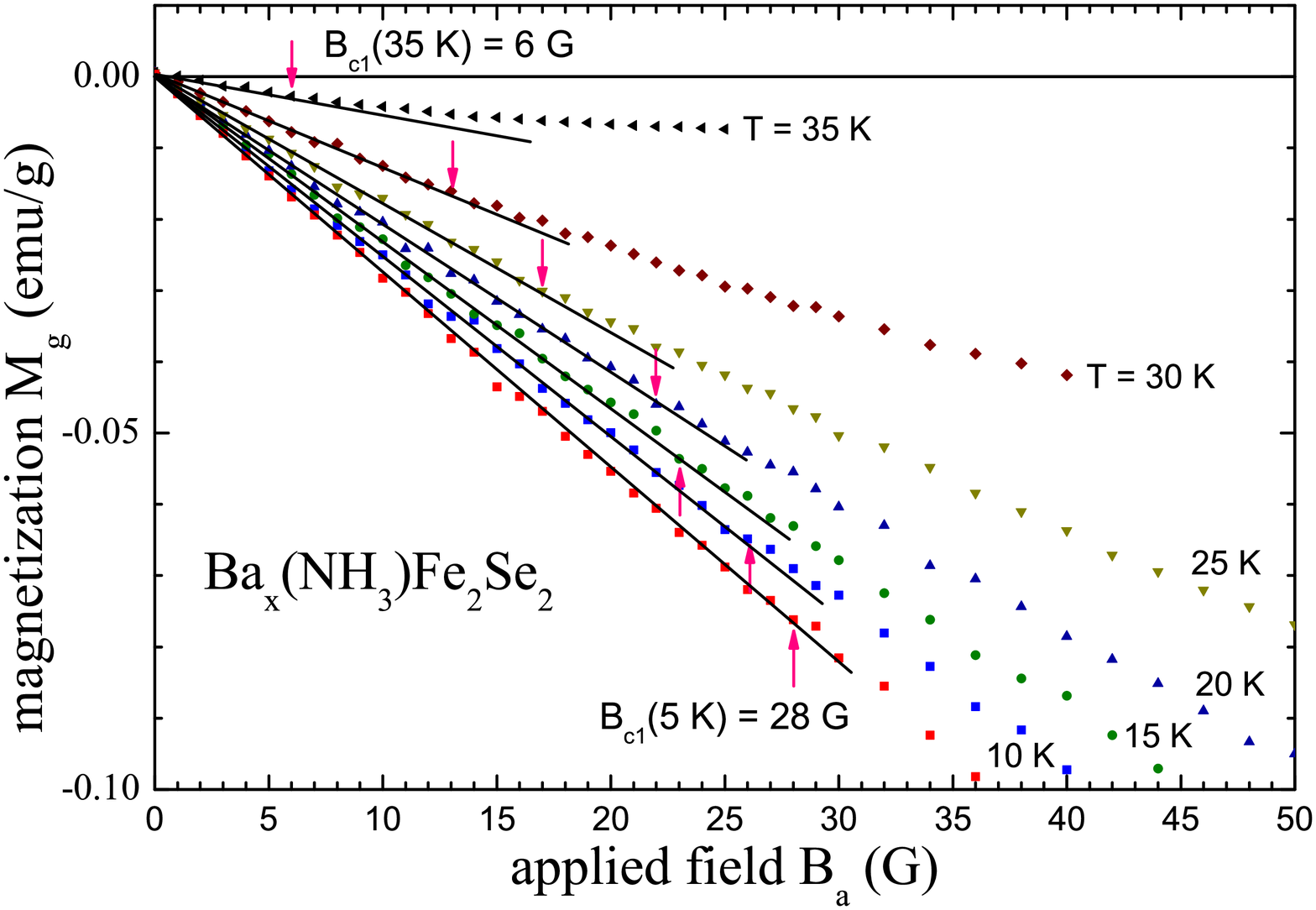}
 \caption{\label{fig.3}
 Low field initial magnetization curves of superconducting $Ba_{x}(NH_{3})Fe_{2}Se_{2}$ at various temperature below $T_{c}$ = 39 K. The curves deviate from liner at $B_{c1}$ as marked by arrows.}
\end{figure}

The initial mass magnetization curves, $M_{g}(T)$, of $Ba_{x}(NH_{3})Fe_{2}Se_{2}$ was shown in figure 3. The lower critical field $B_{c1}(T)$ determined by the deviation point from the low-field linear Meissner response. The obtained $B_{c1}(T)$ decreases monotonically with increasing temperature from 28 G for T = 5 K, to 22 G for 20 K, and then to 6 G for 35 K. Measurements on $Sr_{x}(NH_{3})Fe_{2}Se_{2}$ also revealed a similar behavior of $B_{c1}(T)$ decreased from $B_{c1}$(10 K) = 23 G, to $B_{c1}$(20 K) = 18 G, then to $B_{c1}$(30 K) = 12 G. The observed low $B_{c1}$ values are comparable to those observed for $B_{a} \parallel$ c-axis in $\beta$-FeSe~\cite{Abdel2013}, despite their much higher $T_{c}$'s in these intercalated superconductors. 

\begin{figure}[htb!]
 \includegraphics[width=70mm]{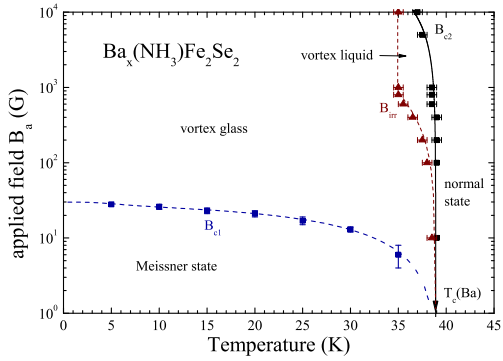}
 \caption{\label{fig.4}
The superconducting $B_{a}$-T phase diagram of $Ba_{x}(NH_{3})Fe_{2}Se_{2}$ derived from magnetic measurements. The $B_{c1}$(0) $\sim$30 G was obtained by simple extrapolation and the linear fitting of  $B_{c2}$(T) was shown by solid curves. The dashed line with $B_{irr}$ is only a guide to the eyes.}
\end{figure}

The superconducting critical fields, $B_{c1}$(T) and $B_{c2}$(T), of $Ba_{x}(NH_{3})Fe_{2}Se_{2}$ obtained from magnetic measurements are summarized in a $B_{a}-$T phase diagram as shown in figure 4. The irreversible line, $B_{irr}$, determined from the deviation points of ZFC and FC curves is also shown as a boundary between vortex glass and liquid states. The zero temperature lower critical field $B_{c1}$(0) $\sim$30 G was easily obtained by extrapolation. An empirical value of $B_{c2}$(0) = 13.4 T was obtained by using the Werthamer-Helfand-Hohenberg formula with a linear slope dB$_{c2}(T_{c})$/dT = -0.497 T/K. The Ginzburg-Landau parameter $\kappa_{Ba}$ = 102 was derived from these extrapolated $B_{c2}$(0) and B$_{c1}$(0) values by solving $B_{c2}$/B$_{c1}$ = 2$\kappa^{2}$/ln$\kappa$, which indicates that $Ba_{x}(NH_{3})Fe_{2}Se_{2}$ is a typical type-II superconductor as expected. Since the superconducting coherent length $\xi_{Ba}(0)$ = 4.96 nm can be easily derived by using the formula $B_{c2}$ = $\Phi_{0}$/2$\pi\xi^{2}$, it is straight forward to calculate the magnetic field penetration depth $\lambda_{Ba}$(0) = $\kappa\xi$(0) = 508 nm. Similar analysis was performed on $Sr_{x}(NH_{3})Fe_{2}Se_{2}$ and the obtained critical fields were $B_{c2}$(0) = 61.4 T and $B_{c1}$(0) = 24 G, by which the superconducting parameters $\kappa_{Sr}$ = 266, $\xi_{Sr}$(0) = 2.33 nm, and $\lambda_{Sr}$(0) = 620 nm were derived.

%dB$_{c2}$/dT = -2 T/K for Sr.
%Despite of the apparent differences of lattice constants, the obtained superconducting parameters between Ba$_{x}$(NH$_{3}$)Fe$_{2}$Se$_{2}$ and Sr$_{x}$(NH$_{3}$)Fe$_{2}$Se$_{2}$ are surprisingly similar that suggests the high-T$_{c}$ of these compounds are mainly determined by the molecule intercalation. This is consistent with the result that ammonia escaping will cause samples to loss superconductivity. The superconducting properties are fine tuned by the insertion of different cations through detailed lattice structure and carrier density variation.

\begin{figure}[htb!]
 \includegraphics[width=70mm]{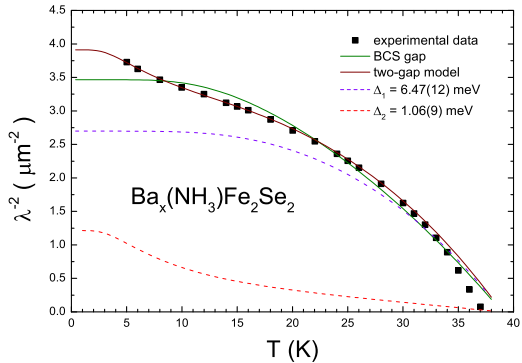}
 \caption{\label{fig.5}
The temperature dependence of $1/\lambda^{2}$, proportional to superfluid density, of $Ba_{x}(NH_{3})Fe_{2}Se_{2}$ estimated from low-field magnetic susceptibility. The data can be well described by a two-gap s-wave model (solid brown curve) with a larger gap $\Delta_{1}$ = 6.47 meV and a smaller gap $\Delta_{2}$ = 1.06 meV. The dashed curves are the individual contributions of each gap. The single-gap BCS behavior is shown in green for comparison.}
\end{figure}

To further investigate superconductivity of the system, temperature dependence of superfluid density, proportional and presented by 1/$\lambda^2$, of $Ba_{x}(NH_{3})Fe_{2}Se_{2}$ was estimated and plotted in figure 5. The London penetration depth $\lambda(T)$ was derived by solving the equation

$\frac{\chi(T)}{\chi_{0}} = \frac{\int(1 - 3 \frac{\lambda}{r} \cot(\frac{r}{\lambda}) + 3 \frac{\lambda^2}{r^2}) r^3 g(r) dr}{\int r^3 g(r) dr}$

where $\chi_0$ is the susceptibility of perfect diamagnetic spheres and g(r) is the grand size distribution function, which was obtained by counting sample powder grands under an optical microscope as $g(r) = 95 \exp(-((\log r + 0.5)/0.382)^{2})$. A constant paramagnetic background was subtracted from the susceptibility for completely counting the diamagnetic signal. 

Since the 10-G applied field became larger than B$_{c1}$ for temperature higher than $\sim$32 K, the penetration depth $\lambda$ values obtained at those temperature were contaminated by vortex formation, thus 1/$\lambda^2$ data for T $>$ 32 K were excluded for further discussion. At the low temperature side, 1/$\lambda^2$ for temperature below 15 K apparently deviated from the saturation behaviour of conventional BCS single-gap model. Referring to similar systems,~\cite{Khasanov2008,Biswas2013} two-gap model was used for analysis and consistent results were obtained.
The temperature dependence of the penetration depth of $Ba_{x}(NH_{3})Fe_{2}Se_{2}$ was fitted by a weakly coupled two-gap s-wave model~\cite{Abdel2013,Biswas2013,Carrington2003} 

$\frac{\lambda^{-2}(T)}{\lambda^{-2}(0)} = \omega \frac{\lambda^{-2}(T, \Delta_{1})}{\lambda^{-2}(0, \Delta_{1})} + (1 - \omega) \frac{\lambda^{-2}(T, \Delta_{2})}{\lambda^{-2}(0, \Delta_{2})}$

where $\lambda(0)$ is the zero temperature penetration depth, $\Delta_{i}$ is the $i$th superconducting gap at T = 0 K and $\omega$ is the weighting factor of the first gap~\cite{Biswas2013}. Each component can be expressed within the local London approximation as

$\frac{\lambda^{-2}(T)}{\lambda^{-2}(0)}$ = 1 + 2$\int_{\Delta_{i}}^\infty()(\frac{\partial f}{\partial E})\frac{EdE}{\sqrt{E^2-\Delta_{i}(T)^2}}$

 where $f = 1/(1 + exp(E/k_{B}T))$ is the Fermi function, and the temperature dependence of the gap is approximated as  $\Delta_{i}(T) = \Delta_{i} \tanh{1.82[1.018(T_{c}/T - 1)]^{0.51}}$.~\cite{Khasanov2008}
The two-gap s-wave model, the brown curve in figure 5, describes the temperature dependence of penetration depth very well. The zero temperature gaps values obtained for $Ba_{x}(NH_{3})Fe_{2}Se_{2}$ are $\Delta_{1}$ = 6.47 meV and $\Delta_{2}$ = 1.06 meV with $\omega$ = 0.69. The derived gap to $T_{c}$ ratios of $2\Delta_{1}/k_{B}T_{c}$ = 3.85 and $2\Delta_{2}/k_{B}T_{c}$ = 0.63 are consistent with those in $Li(C_{2}H_{5}N)_{0.2}Fe_{2}Se_{2}$ ($T_{c}$ = 40 K)~\cite{Biswas2013} and $Li_{0.6}(NH_{3})Fe_{2}Se_{2}$ ($T_{c}$ = 43 K)~\cite{Burrard2013}. 

The slightly bigger value for the $2\Delta_{1}/k_{B}T_{c}$, comparing with the BCS value of $2\Delta_{0}/k_{B}T_{c}$ = 3.35, suggests that the weak-coupling assumption in the two-gap s-wave model used is a good approximation.
Since our samples were randomly orientated powder, the influence of temperature variation of $\lambda_{c}(T)$ should be observed in diamagnetic susceptibility. However, by taking typical anisotropy of iron-selenide systems and the obtained $\lambda(0)$, the estimated value of $\lambda_{c}(0)$ is about 2 $\mu$m which is larger than most ($\sim$ 80$\%$) grain size of the powder. When the c-axes of the single crystal grains have large angles to the applied magnetic field, magnetic field penetrates into the grains completely even at low temperature which makes the effects of temperature dependence are not observed. Thus the estimated $\lambda(0)$ value of $\sim$506 nm could be regarded as the upper bound of $\lambda_{ab}(0)$ and the temperature dependence of 1/$\lambda^2$ represents the supercurrent behaviour in the $Fe_{2}Se_{2}$-layer.   

%These consistent properties suggest the enhancement of these superconductors shared a common mechanism.
%The large gap to the \T_{c} is 3.85 and small gap to the \T_{c} is 0.631. This shows that the large gap is in syrong-coupling limit. 
 %FeSe determined by measuring the magnetic penetration depth using the muon spin rotation technique~\cite{} 
 %FeSe$_{0.85}$ 2$\Delta_{1}$ = 3.85 k$_{B}$T$_{c}$ mev 2$\Delta_{2}$ = 1.07 k$_{B}$T$_{c}$,
%FeTe$_{0.5}$Se$_{0.5}$ 2$\Delta_{1}$ = 4.19 k$_{B}$T$_{c}$ mev 2$\Delta_{2}$ = 1.50 k$_{B}$T$_{c}$,
 %Li$_{0.6}$(NH3)Fe$_{2}$Se$_{2}$  2$\Delta_{1}$ = 5.27 k$_{B}$T$_{c}$ mev 2$\Delta_{2}$ = 0.69 k$_{B}$T$_{c}$ and 
%Li(C$_{2}$H$_{5}$N)$_{0.2}$Fe$_{2}$Se$_{2}$  2$\Delta_{1}$ = 3.96 k$_{B}$T$_{c}$ mev 2$\Delta_{2}$ = 0.57 k$_{B}$T$_{c}$.  
%75.1(3.85) 12.3(0.63) $_{}$ MgB2
% particle size distribution
% chi(T) vs magnetization equation  --  refer to thesis  
% solving two-gap behavior  --  refer to paper
% The temperature dependence of the London penetration depth, $1/\lambda^{2}$, which is proportional to superfluid density, derived from magnetic susceptibility. The data can be well described by a two-gap model (solid brown line) with a larger gap $\Delta_{1}$ = 6.47 meV and a smaller gap $\Delta_{2}$ = 1.06 meV. The saturation yields $\lambda$ = 506 nm. Blue dashed lines are the contributions of the gaps. The green line corresponds to a single-gap BCS curve.

In conclusion, superconducting properties of $A_{x}(NH_{3})Fe_{2}Se_{2}$ (A = Ba or Sr) were studied by magnetic measurements. The high $T_{c}$ after alkali-metal and ammonia molecule intercalation is due to a 3D-like to 2D-like Fermi surface change. Temperature dependence of the London penetration depth was derived from diamagnetic susceptibility for $Ba_{x}(NH_{3})Fe_{2}Se_{2}$ which is well described by a two-gap s-wave model with gap values $\Delta_{1}$ = 6.47 meV and $\Delta_{2}$ = 1.06 meV and $\lambda(0) \sim$ 506 nm.

\ack
This work was supported by the National Science Council of Taiwan under NSC101-2112-M-003-008.

\end{document}